# Ocean Skeletal Structures Hypotheses and Interpretation

**V. A. Rantsev-Kartinov and C.G. Parigger***

*RRC "Kurchatov Institute",
Kurchatov Sq. 1, 123182, Moscow, Russia,
e-mail: rank@nfi.kiae.ru*

**The University of Tennessee / UT Space Institute,*
*411 B.H. Goethert Parkway, Tullahoma, TN 37388, U.S.A.,*
e-mail: cparigge@tennessee.edu
Corresponding Author.

*In this paper we discuss hypotheses on formation of ocean skeletal structures. These structures entered the ocean together with atmospheric precipitation and were assembled from fragments of skeletal structures present in clouds. We base interpretation of this phenomenon on surface tension forces between fundamental tubular blocks of the investigated structures that may also occur beneath the ocean surface. A capillary model is presented to explain formation of a network of interacting tubes. Data about the nature of ocean skeletal structures can be instrumental in modeling many processes associated with physics of the ocean.*

## 1. Introduction

Photographs of the ocean surface that were taken from different altitudes and for varying sea roughness show presence of ocean skeletal structures [1]. The topology of these structures is identical to topology of skeletal structures discovered previously for a wide range of spatial scales, media and phenomena [2-13]. Detailed descriptions and explanations of skeletal properties are given in papers [5,6,8]. Furthermore, hypotheses were suggested based on the primary idea that *only nano-dust components and quantum connections may be responsible for the formation of skeletal structures*. Skeletal structures were first identified in plasma [5]. The reviews [7, 10] indicate a role of the nano-dust in formation of the skeletal structures over a range of length scales that spans 30 orders of magnitude.

## 2. Hypotheses

The basic hypotheses about the formation of ocean skeletal structures were suggested in Ref. [1]. Contents of these hypotheses included: **a**) Particles of dust which are capabale to generate structures are released into the atmosphere by active volcanoes or by dusty storms in deserts. Subsequently, these particles make up the ocean skeletal structures. Analogous to previous descriptions so-called carbon-nano-tubes (CNT) and nano-particles of other chemical elements were considered as the basic element in the skeletal structure [2-13]; **b**) Earth electrical field and atmospheric electricity can cause formation of skeleton structures of clouds whose fragments come in contact with the ocean surface due to different atmospheric phenomena; **c**) The skeleton structures show an active surface and residual buoyancy. They float on the ocean surface due to adsorbing air that is dissolved in water on their own surface and due to partial filling by sea foam; **d**) The skeleton structures are comprised of three different phases, namely hard-solid, fluid and gaseous, thereby creating the possibility of surface tension even under the ocean surface by connecting separate building blocks into a united network; **e**) The strength of the oceanic structure and its separate blocks is determined by the packing compactness of the skeleton structure; **f**) The skeletal properties of the ocean skeletal structures become clear as surface tension induces flexible joints within the ocean skeletal network, analogous to joints in a skeleton.

## 3. Oceanic Skeleton Structure Observations

According to Ref. [1] only several block types of ocean skeletal structures can be recognized in image analysis using the so-called multilevel-dynamical contrast (MDC) method of image processing [10]. Also, an increase of the fundamental structure size is correlated with an increase in roughness of the ocean surface. In this work, we present an estimate for durability and buoyancy of the ocean skeleton structure in terms of vertically floating cylinders (VFCs). These cylinders are coaxially-tubular, compactly packed blocks that consist of previously generated structures that came into contact with the ocean surface during storms. Some structures may be partially destroyed during heavy weather. The cylinders contain smaller size cylinders that may be as small as capillaries with diameters of tens of microns, in a form similar to CNTs mentioned above. Such capillaries fill the blocks of the ocean skeletal structure. Composition of individual segments of the skeleton structure may vary. However, this variation does not affect durability. For example, we will consider the coaxially-tubular element with wheel-like structure on its ends and with parallel knitted-needles directed along the lateral, cylindrical surface. This particular block is relatively strong and stable with respect to conserving its shape.

The strongest elements of the fundamental building blocks may be constructed with capillaries. These capillaries may be partially or completely filled by sea foam, thereby providing buoyancy. Figure 1 illustrates this idea: It shows a fragment of the ocean surface imaged from an altitude of 150 m during the hurricane Belle [14]. The image width corresponds to approximately 15 m. One can demarcate almost half of a vertically oriented floating cylinder of diameter 25 m. The diameter of the central, axial tube amounts to 3 m with a diameter of a dark ring of 1.5 m. The structure is approximately 1.5 m above the local sea level. OO' indicated the direction of the cylinder axis. An angle between the OO' direction and vertical line is about 30°. In order to clearly show the cylindrical structure we applied the multilevel dynamical contrast (MDC) method developed and described earlier by one of us (VAR-K) for analysis of barely observable structures in images of various phenomena and environments [3,4,10].

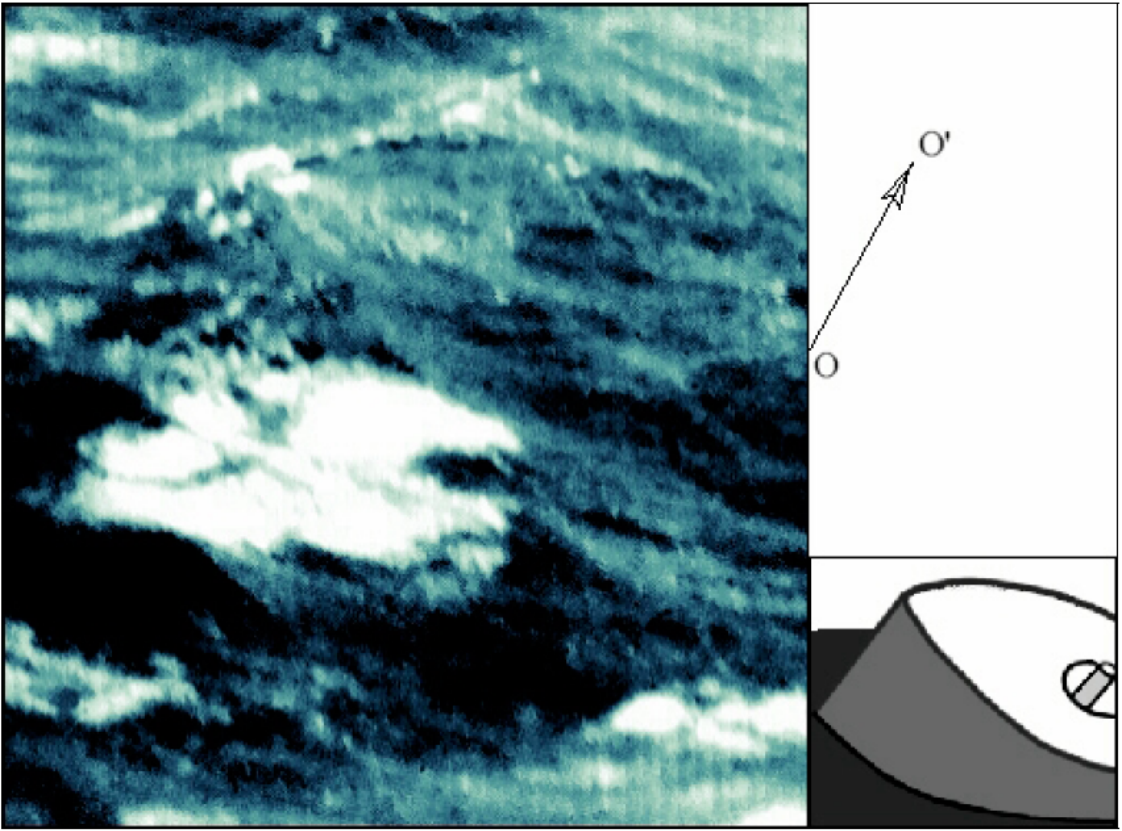

**Figure 1:** Left: Section of the sea surface image taken during the Hurricane Belle from 150 m altitude, see *National Oceanic and Atmospheric Administration* (NOAA) collection [14]; Right: Schematic of the cylindrical structure. Reproduced with permission from Ref. [1].

**4. Interpretation of the Observable Phenomena**

The hypothesis **d** mentioned in Section 2 suggests that *presence of different phases may introduce new mechanisms for skeleton structure formation* [7]. Important in this context will be the question: "Are there sufficient surface tension forces and capillary phenomena that can explain the observed vertically-floating cylinders, which are filled up with a sea water and/or foam and which extend up to 1 meter above the sea level?" We consider (i) that during a storm forces may occur that can elevate a cylinder (assembled of thin capillaries) to sufficient height above sea level, and (ii) that capillary walls are moist. For this scenario we will estimate the height from the capillary action. If the vertically-floating cylinder (VFC) construction is filled with closely packed thin capillaries (oriented parallel to the VFC lateral surface) and the water weight of its column is balanced by surface tension forces in the capillaries, then the VFC weight is balanced by forces responsible for the structure's buoyancy. In an idealized block consisting of smaller sized capillaries that fill up of the VFC, the water weight that is lifted above sea-level is balanced by the capillary forces. The block will keep its form, and its end surface will be swept by water due to the rough ocean surface as a solid / open working surface. Consequently, as the VFC volume partially fills up with sea foam from a rough ocean surface, it will float above the mean sea level and at a specific height. The capillaries forces sustain the water in its canals. The VFC floats and conserves its own geometrical form with the vertical oriented side surface until sea foam enters the capillary canals. Sea foam is generated by overturning waves during storms that may occur in the open ocean. Along the vertical direction the VFC is stretched by its own weight with the top end floating above the average sea-level. In the radial direction the VFC structure shows strength characteristic of a cylindrical lock comprised of parallel capillaries connected by surface tension forces. One can show for the breaking point $E$ of the vertically floating cylinder (VFC)

$$E \approx 4 \times 10^{-3} \sigma/d \ [g/cm^2]. \qquad (1)$$

Here, $\sigma$ is the surface-tension coefficient of the water [dynes/cm] and $d$ is the capillary diameter [cm], which fill up the VFC body. (The surface tension of water at 25 degree Celsius amounts to 72 dynes/cm.) The thin capillaries strengthen the VFC construction. When the VFC axis is deflected by an angle $\alpha$ relative to the vertical direction, then the capillary length filled up by the water amounts to

$$l \propto 1/\cos\alpha . \qquad (2)$$

Fluid may escape from the top of the capillary causing the inclined block to marginally sink and to incrementally lose stability. This loss of stability may be accompanied by destruction of the inclined lateral surface that may be observed.

The maximum sizes of diameters of capillaries that confine the VFC is estimated in the following. As water is lifted by the capillaries up to height $h$ the diameter satisfies the inequality

$$d \leq 4\sigma/gh\rho . \qquad (3)$$

Here, $g$ is the gravity acceleration constant, and $\rho$ is the water density. Using tabulated values one can find the maximum size of the capillary diameter. For a height of $h \sim 1$ m, the capillary diameter amounts to

$$d \leq 30 \; \mu m. \tag{4}$$

For large enough buoyancy (in order to hold a water column, which is lifted up to some height $h$ over the mean sea level due to the capillary forces) the capillary construction will float and conserve its cylindrical shape with the vertically oriented side surface.

In the following we estimate the buoyancy of the vertically floating cylinder (VFC). Also, we estimate its ability to carry the water weight with a diameter equal to an external diameter of a floating structure for a height $h$ above the mean sea level. At the outset it is necessary to generate a model for the construction of self-similar tubular structures. Here we consider a one/two-level carbon nano-tube (CNT) as the basic element of our structure, 10-nm in size. Such a CNT can be found with a large probability due to an energetic interaction with crystalline carbon. The carbon crystalline plane is a mosaic with flat sections (with typical size $\sim$ 10-nm and compactly arranged carbon hexahedrons of side $\sim$ 1.42 Å) with relaxed connections. The first generation CNT is rolled up into a tube-like scroll containing approximately $\sim 8 \times 10^3$ carbon atoms, and the mass amounts to $m_1 \sim 1.7 \times 10^{-22}$ kg. Subsequently, one arranges a square from first-generation CNTs such that they form the same hexahedrons but with side of $\sim$ 10-nm, followed by rolling it into a tube similar as before, labeling it the second generation. Subsequent generations are formed similarly. The mass, length and diameter of n-th generation tubes is determined as follows:

$$m_n \sim 1.7 \times 10^{-22} \times 10^{4(n-1)} \; kg, \quad l_n \sim 1.42 \times 10^{-10} \times (75)^7 \; m$$
$$\text{and} \; d_n \sim 3 \times 10^{-9} \times (75)^{n-1} \; m. \tag{5}$$

From Equation (5) one finds for Earth conditions that one should be able to observe wheel-like structures of 9-th generation, i.e., $d_9 = 3 \cdot 10^3 \; km$. Yet the maximum size of the wheel-like diameter amounts to approximately 650 km, as inferred from analyses of a data base of ocean surface images captured from satellite. On the ocean surface one may observe full-length coaxially-tubular structures (see Ref. [10]) in shapes of a vertically floating cylinder (VFC) up to 7-th generation. Indeed, generation $n = 7$ implies a diameter $d_7 \sim 530\,m$ and length $l_7 \sim 2\,km$. In that way, the coaxially-tubular structure of a 7-th generation VFC can freely move along the ocean surface for sufficient ocean depth. Similar structures of the 8-th generation show a diameter $d_8 \sim 40\,km$ and length $l_8 \sim 140\,km$. This structure may only be seen as horizontally floating cylinder (HFC) or as shortened VFC. The ends of the CNTs up to 3-rd generation may become clogged (providing thereby the buoyancy for the ocean skeletal structure) with a flat phytoplankton, which shows a size of about

$$\delta \sim 5\,\mu\text{m}. \tag{6}$$

Between this length and the magnitude of the capillary diameter, capable of lifting the water column to height $\sim 1$ m, there is a gap of one order magnitude. CNTs can effectively absorb air dissolved in the sea water on their surface and (according to formation and gathering of air bubbles on its surface) create some additional buoyancy for the ocean skeletal structure blocks. One finds that only 3-rd generation CNTs can lift the water column by one meter. Indeed,

$$d_3 \sim 17\,\mu\text{m}, \tag{7}$$

which is consistent with Equation (4). Moreover, these CNTs are not populated by phytoplankton particles since these are 3 to 4 times smaller (see Equ. (6)). The specific gravity selected for the ocean skeletal structure, for example of diameter $\sim 8$ m, is determined by its generation number. This generation number can be determined from Equ. (5), namely for n $\sim 6$ one finds $d_6 = 7$ m. From here one calculates a density of the initial tube skeleton $\sim 1.3 \times 10^{-5}\,\rho$ – a magnitude that may be ignored in our analysis. During a strong storm the initial skeleton structure is not strong enough to hold out against high sea. The skeleton breaks into pieces and forms a stronger structure, as can be inferred from analyses of ocean surface images captured during heavy storms. The new structure is a tubular structure with tubes corresponding to the determined generations, but they are compactly filled by tubes of smaller diameter (another boundary case). However, some of the inner tubes may remain unfilled.

In the following we describe modifications of the initial skeletal structure in the sea water and how it transforms into the ocean skeleton structure. As indicated above, we need to focus only on 3-rd generation tubes that will build the VFCs. Strong roughness of the ocean surface destroys the initial skeleton structure. Subsequently, the ocean surface will be covered by these tubes, viz. capillaries. For strong wind the waves have a primary orientation which causes the tubes to line up along the wave crest. The tubes become interconnected due to the surface tension, the structure increases and in turn cover the ocean surface as a connected, compact carpet. The images analysis shows connected similar filaments (and even including considerably larger diameter filaments) that are joined to multiples of individual filaments lengths up to one hundred meters. Intense, rough ocean surface induces rolling up of this "carpet" into cylindrical tubes of compactly packed tubes of smaller diameter. Simultaneously, capillaries fill in residual blocks that originate from destruction the initial skeleton structure.

Statistics of the investigated topology of the observed tubes can give (in zero order) a method for its construction – the cylinder of present generation is a sheath of cylinders from a previous generation and (most likely) six of the same cylinders around it, which are contained in a single cylindrical shell. Moreover, an unfilled space may occur which may or may not be filled by cylinders (capillaries) of earlier generations. The total filling percent may be almost 100 %. The mentioned construction method for the generations appear quite natural, because the process of self-assembly and enlargement of cylinders proceed in tandem with intense waves and increased power. Moreover, strong roughness of the ocean surface and overturning of the waves causes formation of sea foam. The sea foam increases the buoyancy of the skeletal structures (which form during that process) as it promotes decrease of mean specific gravity of water, which in turn fills the blocks of the ocean skeletal structure. Let us estimate a specific gravity of carbon for an ideal VFC that is completely packed with 3-rd generation CNTs. We find a specific gravity of approximately $1.3 \times 10^{-3}$ $\rho$, which is 100 × larger than the density of the condensed component of the initially considered ideal model for the skeleton structure. As the ocean surface roughness increases, the specific gravity of the dusty component of the ocean skeleton structure increases near the water surface. But it makes an insignificant contribution to the mean specific gravity of sea water, because salt content and other impurities contribute almost 20 times as much than the dusty component. It is noteworthy, that only by sampling water inside ideal VFC's and by sampling water during a hurricane may result in high micro-dust content of the sea water. For all other cases its content is typically in the order of ~ $1.3 \times 10^{-5}$ $\rho$. For such cases, the tubular structure in the water is 'invisible' or is in general not seen. Therefore, it is very problematical to show its presence in the water by means of simple evaporation and separation. Presence of tubular structure in the ocean water can only be evaluated by means of a delicate physical/chemical analysis. Equally, skeleton structures can rely on analysis of particular properties and phenomena, which may become apparent at different stages of ocean surface roughness that occur as anomalous atmospheric phenomena that are connected with the sea.

Let us now estimate the buoyancy of the VFC with the maximal diameter, which was observed as described above and that towers above sea-level during some finite time. Of course, implied is the assumption that we have convinced ourselves of the existence of the VFC and took this fact as granted. From observations during strong storms one can infer that similar horizontally floating cylinders (HFC) have a length L in the order of 50 to 100 m. It is already sufficient to estimate (for on-average similar VFC volume) the specific gravity of the filling water in order to determine whether the structure will float at sea level. For the water-filled VFC with mean specific density $\rho$, floating above the mean sea level at height $h$ of approximately 1 meter, one may write for the ratio

$$\rho/\rho_a = (L-h)/L . \qquad (8)$$

Inserting the numerical values for $L$ and $h$, we find

$$\rho_a = (0.98 - 0.99) \times \rho . \qquad (9)$$

That is, the total volume of air in the water (which fills the VFC) consists only (1-2) %. During escape of air bubbles into the atmosphere and sea foam from the capillaries, the VFC will sink incrementally down to the water surface. Such anomalous VFC buoyancy exists due to differences of two kinds – first, due to escape of air bubbles from the open water and, second, due to escape from the VFC capillaries. The VFC is lowered to the water surface, when the specific gravity of the water in the capillaries equals the specific gravity of the open water.

Higher concentrations of bubbles appear when the crest of a wave overturns on the VFC or when the sea foam together with the water flows into it from below. The latter takes place when power streams of foamed water almost entirely fill up the VFC. The VFC will float when filled with plenty of foam. Thus, *the CNT component provides an action from the surface tension force even under the ocean surface created in a way to guarantee sustenance of the created ocean skeletal structures including their buoyancy*. The total force of the VFC cohesion may be very considerable (due to a developed lateral surface of the block) since it is compactly filled by thin interacting capillaries. Since the water level in the capillaries is completely determined (see Equ. (3)) there are no forces, which potentially can cut-off the VFC near its base. For a VFC that is assembled of 5-th generation coaxially-tubular blocks we estimate the minimum angle $\alpha_{min}$ of the VFC declination from the vertical line to be

$$\alpha_{min} \approx \arcsin\left(2\sigma/d\rho g\right) \quad (10)$$

The interior of the 5-th generation VFC is filled with closely-packed blocks of a smaller size down to thin capillaries of tens of microns in size, for example, in the form of 3-rd generation coaxial tubes with diameter $d_3$. For diameters $d = d_5 \sim 10\ cm$ (see Equ. (5)), we find $\alpha_{min} \sim 1.5 \times 10^{-2}$, in other words such inclined VFC will fall to the water surface because the 5-th generation coaxial tubes near the upper end will tear off from cylinder and bend over the water surface on the side with a negative inclination. The VFC illustrated in Figure 1 shows an angle of inclination relative to the vertical line of about 30° and it sustains the geometrical form. From this it follows (see Equ. (10)) that

$$d_{max} \leq 4\sigma/\rho g = 3000\ \mu m, \quad (11)$$

which almost corresponds to $d_4 \sim 1300$ μm. Thus, from observations we infer that

$$\delta < d_3 \approx d \leq d_4, \text{ in other words, } 5\mu m < d \leq 3000 \mu m. \quad (12)$$

During high levels of ocean surface roughness its strips uncover the HFC structures which show lengths of typically 50 to 100 m. From this we conclude the aspect ratio of $D/l \sim 3 \cdot 10^{-4}$. During mean levels of ocean surface roughness the HFC sometimes overlap the distance between two neighboring waves, viz. similar to a convex bridge. In such case we can estimate a mean value of the structure strength.

Figure 2 shows a section of the mentioned structure during mean levels of ocean surface roughness. The image width corresponds to 1.7 m. At this magnification the ocean skeletal structure appears as long and nearly straight-line dendrite construction with branches that are immersed to some depth.

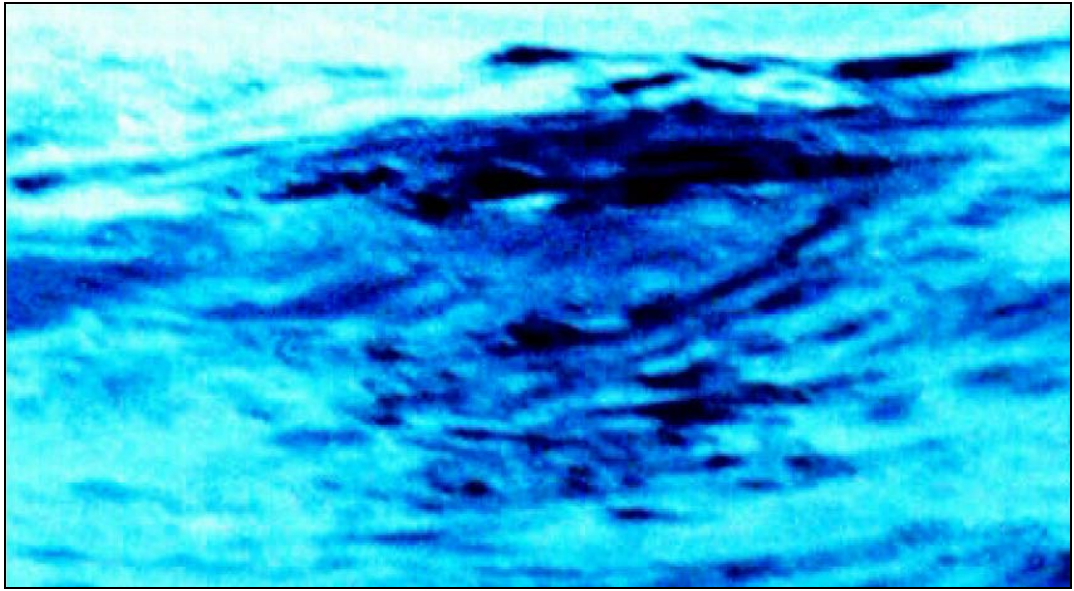

**Figure 2:** Horizontally floating cylinder (HFC) connects two neighboring waves similar to a convex bridge.

In reference to Figure 2, suppose the weight of the indicated beam is determined by the water specific gravity, as water fills the beam. As we take into account the fragment scale mentioned above, we can determine the average value of the breaking point E. One finds that $E \geq 100 g/cm^2$ ($D \sim 10\ cm$, and $l \sim 10^2\ cm$). We assume that such structure is comprised of a bunch of long capillaries, connected to each other due to action of surface tension forces. Consequently, we can estimate the upper value of the capillary diameters to be $d \leq 30\ \mu m$. This value corresponds to the value for VFC's given in Equ. (7).

## 5. Conclusions

Key elements of ocean skeletal structures are dynamics of formation and effects of ocean storms. Horizontally floating cylinders may float marginally close to the ocean surface. Rise above the mean sea-level however will cause that parts of this structure will dry up due to the Sun's heat, and consequently connecting fluids will be lost and disintegration into fragments will occur. These fragments may float or immerse into the water depending on their individual buoyancy. The floating fragments tend to coalesce into a new ocean skeletal structure. The formation time should be an individual and finite quantity for a more or less stable and homogeneous structure corresponding to roughness of the local ocean surface. The individuality of the ocean skeletal structure in a given region is however determined by prior events that occurred during a wide time interval. For example, consider ship propellers that modify and grind the ocean skeletal structure disturbing homogeneity: It may take several hours over spatial scales of several 100 km to form and/or re-form a new, homogeneous ocean skeletal structure. This can be optically measured: Reflectivity of an ocean surface depends on presence, form and scales of the ocean skeletal structure.

## Acknowledgements

One of us (VAR-K) thanks A.B. Kukushkin for his decade-long collaboration. Special thanks go to V.I. Kogan for invariable support and interest in research of skeletal fractal structures. CGP acknowledges support in part from the Russian Academy of Natural Sciences and in part from the Center for Laser Application to attend the 2010 Plasma Science conference in Zvenigorod, Russia, that allowed us to commence international collaboration on this subject.

## References

1. V.A. Rantsev-Kartinov, "Evidences for Skeletal Structures in the Ocean", Phys. Lett. A **334**: 234-242, 2004.

2. V.A. Rantsev-Kartinov, http://www.arxiv.org/ftp/physics/papers/0401/0401139.pdf

3. A.B. Kukushkin and V.A. Rantsev-Kartinov, "Dense Z-Pinch Plasma as a Dynamical Percolating Network", *Laser and Particle Beams* **16**: 445-471, 1998.

4. A.B. Kukushkin and V.A. Rantsev-Kartinov, "Self-similarity of Plasma Networking in a Broad Range of Length Scales: from Laboratory to Cosmic Plasmas," Rev. Sci. Instrum. **70**: 1387-1391, 1999.

5. A.B. Kukushkin and V.A. Rantsev-Kartinov, "Filamentation and networking of electric currents in dense **Z**-pinch plasmas", Fusion Energy 1998 / Proc. 17th IAEA Fusion Energy Conference. Yokohama, Japan, 19-24 Oct 1998 / Vienna: IAEA-CSP-1/P **3**: 1131-1134, 1999.

6. A.B. Kukushkin and V.A. Rantsev-Kartinov, "Microsolid tubular skeleton of long-living filaments of electric current in laboratory and space plasmas," Proc. 26-th Eur. Phys. Soc. conf. on Plasma Phys. and Contr. Fusion / Maastricht, Netherlands: 873-876, 1999.

7. A. B. Kukushkin and V. A. Rantsev-Kartinov, "Similarity of skeletal objects in the range $10^{-5}$ cm to $10^{23}$ cm," Phys. Lett. **A 306**: 175-183, 2002.

8. A. B. Kukushkin and V. A. Rantsev-Kartinov, "Long-living filamentation and networking of electric current in laboratory and cosmic plasmas: from microscopic mechanism to self-similarity of structuring", Current Trends in International Fusion Research: Review and Assessment (Proc. 3rd Symposium, Washington D.C., March 1999), Ed. E. Panarella, NRC Research Press, Ottawa, Canada, 121-148, (2002);

9. A. B. Kukushkin and V. A. Rantsev-Kartinov, "Skeletal structures in high-current electric discharges and laser-produced plasmas: observations and hypotheses"Advances in Plasma Phys. Research, 2002, vol. 2 (Ed. F. Gerard, Nova Science Publishers, New York), pp. 1-22.

10. A.B. Kukushkin and V.A. Rantsev-Kartinov, "Universal Skeletal Structures in Laboratory and in Space," Science in Russia **1**: 42-47, 2004.

11. B. N. Kolbasov, A. B. Kukushkin, V. A. Rantsev-Kartinov, and P.V. Romanov, " "Similarity of Micro- and Macrotubules in Tokamak Dust and Plasma", Phys. Lett*.* A, 2000, vol. 269, pp. 363-367.

12. B. N. Kolbasov, A. B. Kukushkin, V. A. Rantsev-Kartinov, and P.V. Romanov, "Skeletal dendritic structure of dust microparticles and of their agglomerates in tokamak T-10", Phys. Lett. **A,** 2001, vol. 291, pp. 447-452.

13. B. N. Kolbasov, A. B. Kukushkin, V. A. Rantsev-Kartinov, and P.V. Romanov, "Tubular structures in various dust deposits in tokamak T-10," Plasma Devices and Operations **8**: 257-268, 2001.

14. http://www.photolib.noaa.gov/flight/images/big/fly00164.jpg